\documentclass[10pt,twocolumn,floatfix,superscriptaddress,amsmath,showpacs,aps,pre]{revtex4-1}

\begin{document}
\title{Nonlinear inhomogeneous Fokker-Planck equation within a generalized Stratonovich prescription}
\author{Zochil Gonz\'alez Arenas}
\affiliation{Centro Brasileiro de Pesquisas F{\'i}sicas\\ National Institute of Science and Technology for Complex Systems, 
Rua Xavier Sigaud 150, 22290-180,Rio de Janeiro, RJ, Brazil}
\author{Daniel G.\ Barci}
\affiliation{Departamento de F{\'i}sica Te\'orica,
Universidade do Estado do Rio de Janeiro, Rua S\~ao Francisco Xavier 524, 20550-013,  Rio de Janeiro, RJ, Brazil}
\author{Constantino Tsallis}
\affiliation{Centro Brasileiro de Pesquisas F{\'i}sicas\\ National Institute of Science and Technology for Complex Systems, 
Rua Xavier Sigaud 150, 22290-180,Rio de Janeiro, RJ, Brazil}
\date{\today}

\begin{abstract}
We deduce a  nonlinear and inhomogeneous Fokker-Planck equation within a  genera\-li\-zed Stra\-tono\-vich, or 
stochastic $\alpha$-, prescription ($\alpha=0$, $1/2$ and $1$ respectively correspond to the It\^o, Stratonovich 
and anti-It\^o prescriptions). We obtain its stationary state $p_{st}(x)$ for a class of constitutive relations 
between drift and diffusion and show that it has a $q$-exponential form, $p_{st}(x) = N_q[1 - (1-q)\beta V(x)]^{1/(1-q)}$, with an index $q$ which does {\it not} depend on $\alpha$ in the presence 
of any nonvanishing nonlinearity. This is in contrast with the linear case, for which the index $q$ is $\alpha$-dependent.
\end{abstract}

\maketitle

\section{Introduction}

The nonlinear Fokker-Planck (FP) equation has been largely used to study a wide class of physical systems which exhibit 
anomalous diffusion~\cite{Plastino95,Tsallis96,Frankbook,Tsallis2002,CuradoNobreetal}. A particular feature of this equation is that its stationary solutions 
are  probability distributions obeying nonextensive statistical mechanics~\cite{Tsallis88,books}.

From a mesoscopic point of view, nonlinear FP equations are 
related with a class of Langevin equations with multiplicative noise~\cite{Lisa1}.  In these processes, the inhomogeneity of the 
diffusion function is proportional to a function of the probability density itself. Therefore, the computation of stochastic 
trajectories turn  out to be very  cumbersome. Indeed,  one needs, for each noise realization, to know the complete time evolution of the probability density, making the problem a self-consistent one, very hard to deal with. Moreover, it is well known that, to correctly define the stochastic multiplicative 
process it is necessary to fix a  prescription to perform the Wiener integrals. The stochastic evolution depends on 
this prescription and the final stationary state, if it exists, might also be prescription-dependent. 
The most popular conventions are the It\^o and Stratonovich ones. However, it is possible to work within a more general 
scheme usually referred to  as \emph{generalized Stratonovich prescription}~\cite{Hanggi1978} 
or \emph{$\alpha$-prescription}~\cite{Janssen-RG}. This convention is parametrized by a parameter $\alpha$
($0 \le \alpha \le 1$) and  recovers the It\^o and Stratonovich prescriptions as the $\alpha=0$ and $\alpha=1/2$ particular cases.
The concept of equilibrium in these systems should be carefully defined, since the forward and backward stochastic 
evolutions are generally performed with different dual prescriptions and, as a consequence,
 the usual detailed balance relations should be properly generalized~\cite{ArenasBarci3}. 

Recently, a class of inhomogeneous and nonlinear FP equations~\cite{MarizTsallis} was considered within the It\^o prescription. It was shown that
stationary solutions for the probability density are of the $q$-exponential form, namely
\begin{eqnarray}
p_{\rm st}(x) = P_q(V) &\equiv& N_q \,e_q^{-\beta V(x)}  \nonumber \\
&\equiv& N_q \, [1 - (1-q)\beta V(x)]^{\frac{1}{1-q}},
\label{q-exp}
\end{eqnarray}
where $V(x)$ is any confining potential, $N_q$ a normalization constant, $\beta$ and $q$ are two real numbers 
characterizing the distribution, and $e_1^z=e^z$; 
$\beta$ is an inverse effective ``temperature'' \cite{effectivetemperature}. In particular, 
$\lim_{q\to 1} P_q(V)= N_1\, e^{-\beta V(x)}$, the usual Boltzmann-Gibbs (BG) distribution. 

As mentioned above, in \cite{MarizTsallis}, the FP equation was deduced within the It\^o convention.  In the present paper 
we analyze how the stationary distributions depend on the particular prescription used to define the stochastic process.
To do this, we first deduce the  inhomogeneous and nonlinear FP equation within the generalized Stratonovich convention, 
parametrized by $\alpha$, and then we look for stationary solutions of a family of processes,  defined by a particular class of constitutive relations 
between drift and dissipation.  As we shall see, the model is parametrized with two real numbers, namely $\eta$ and $\theta$, defined hereafter. The first one measures
the nonlinearity of the system,  while the second one is related with inhomogeneity;  
the point $(\eta,\theta)=(0,0)$ corresponds to the linear homogeneous particular case and represents a normal diffusion process. The stationary-state solutions depend on the 
values of these parameters. In particular, it will become clear that the solutions are,  in the linear  limit $\eta \to 0$, nonanalytic in the space $(\eta, \theta)$.  We have found that, for the general case in which $(\eta,\theta)\neq (0,0)$, the stationary 
probability distributions
are $q$-exponentials with an index $q$ which is {\em independent of the stochastic prescription}. The different conventions, characterized by $\alpha$, do  modify the temperature parameter $\beta$, but {\it not} $q$.

The paper is organized as follows. In section \ref{NLFP} we present the nonlinear  inhomogeneous FP equation and
define our model. In section \ref{eta0} we study the linear limit ($\eta = 0$, $\forall \theta$), and in section \ref{etaneq0} we address the general case.
Finally, we discuss our results in section \ref{conclusions}.

\section{The nonlinear Fokker-Planck equation for multiplicative Markov processes}
\label{NLFP}
Consider a Markovian multiplicative stochastic process described by the Langevin equation
\begin{equation}
\frac{dx}{dt} = F(x,t) + [\phi(x,t)]^{1/2}\xi(t),
\label{Langevin}
\end{equation}
where $\langle\xi(t)\rangle=0$ and $\langle\xi(t)\xi(t')\rangle=\delta(t-t')$.  $F(x,t)$ is the drift force, and $\phi(x,t)$ is in principle an arbitrary function that models the state-dependent 
diffusion process.   As it is well known, this equation should be complemented with a prescription to integrate 
the Wiener integral. In this paper, we use the  \emph{generalized Stratonovich prescription}~\cite{Hanggi1978} 
or \emph{$\alpha$-prescription}~\cite{Janssen-RG}. Briefly speaking, it
is necessary to give sense to the ill-defined product
$[\phi(x(t),t)]^{1/2}\xi(t)$, since $\xi(t)$ is delta-correlated. 
By definition, the Riemann-Stieltjes integral of a Wiener process $W(t)$ with 
$\xi(t)=dW(t)/dt$  is
\begin{eqnarray}
 \lefteqn{
\int   [\phi(x(t),t)]^{1/2}\;  dW(t)=} \nonumber \\
&& \lim _{n\to\infty}
\sum_{j=1}^n 
[\phi(x(\tau_j),\tau_j)]^{1/2}(W(t_{j+1})-W(t_j))
\label{eq.Wiener}
\end{eqnarray}
where $\tau_j$ is taken in the interval $[t_j,t_{j+1}]$ and the limit is taken
in the sense of {\em mean-square limit}~\cite{Gardiner}. For a smooth measure
$W(t)$, the limit converges to a unique value, regardless the value of
$\tau_j$. 
However, $W(t)$ is not smooth,  in fact, it is nowhere integrable. In any
interval, white noise fluctuates an infinite number of times with infinite
variance. 
Therefore, the value of the integral depends on the prescription for the choice
of $\tau_j$.
In the ``generalized
Stratonovich prescription''  we choose 
\begin{equation}
x(\tau_j)=(1-\alpha)x(t_j)+\alpha x(t_{j+1})\mbox{~~ with~~} 0\le \alpha
\le 1. 
\label{eq.prescription}
\end{equation}
In this way, $\alpha=0$ corresponds with the pre-point It\^o interpretation and
$\alpha=1/2$ coincides with the (mid-point) Stratonovich one.  
Moreover, the post-point prescription, $\alpha=1$, is also known as the kinetic
or anti-It\^o 
interpretation.
In principle, each particular choice of $\alpha$ fixes a different stochastic
evolution. 

In many physical applications, a weakly colored Gau\-ssian-Mar\-kov noise with a
finite variance~\cite{Hanggi-shot} is considered. 
In this case, there is no problem with the interpretation of
equation~(\ref{Langevin}) and the limit of infinite variance can be taken at
the end of the calculations. 
This regularization procedure is equivalent to the Stratonovich interpretation,
$\alpha=1/2$~\cite{vanKampen,Zinn-Justin}. However, in other applications, 
like chemical Langevin equations~\cite{vanKampen} or econometric
problems~\cite{Mantegna,Bouchaud},  
the noise can be considered principally white, since it could be a reduction of
jump-like or Poisson-like processes. 
In such cases, the It\^o interpretation ($\alpha=0$)
should be more suitable. Hence, the interpretation of
equation~(\ref{Langevin}) depends on the physics behind 
a particular application. 
Once the interpretation is fixed, the stochastic dynamics is unambiguously
defined. 

From the stochastic equation~(\ref{Langevin}) it is possible to derive a Fokker-Planck equation,
given by~\cite{Celia,Coutinho,ArenasBarci2,ArenasBarci3}
\begin{eqnarray}
\frac{\partial p(x,t)}{\partial t} &=& - \frac{\partial}{\partial x} \left\{\left[ F(x) 
+  \frac{\alpha}{2} \frac{\partial \phi(x,t)}{\partial x}  \right] p(x,t)\right\}  \nonumber \\
&& + \frac{1}{2} \frac{\partial^2}{\partial x^2}\left\{\phi(x,t)p(x,t)\right\},
\label{FP} 
\end{eqnarray}
where $p(x,t)$ is the time-dependent probability distribution and $\alpha \in
[0,1]$ parametrize the stochastic prescription. 

If the function $\phi(x,t)$ is an  ``external'' fixed  function, modeling  a simple state diffusion process, 
then, Eq.~(\ref{FP}) is linear. However, as discussed in Ref. ~\cite{Lisa1}, 
the diffusion function could depend on the probability distribution itself,
for instance,
\begin{equation}
\phi (x,t) = D \left[ g(x)\right]^{\theta} \left[ p(x,t)\right]^{\eta},
\label{phi}
\end{equation}
where $D$ is a constant diffusion coefficient, $g(x)$ is an arbitrary well-behaved function and $p(x,t)$ 
is a solution of the FP equation. With this choice, Eq.~(\ref{FP}) is a nonlinear equation describing 
a state-dependent diffusion process with non-trivial particle-bath couplings~\cite{Lisa1}. 
The real constants $\theta$  and $\eta$  control the relative strength of these
effects. 
For instance, the point $\eta=\theta=0$ represents a normal
diffusion process driven by a stochastic additive Langevin equation. On
the other hand, the line $\eta=0, \theta\neq 0$, 
represents a usual state-dependent diffusion process, described by a multiplicative Langevin equation. 
Moreover, the general case $\eta\neq 0$, is a multiplicative process, whose
diffusion functions should be self-consistently computed by solving the related
nonlinear FP equation. 

Eq.~\eqref{FP} can be written as a continuity equation,
\begin{equation}
\frac{\partial p(x,t)}{\partial t} = \frac{\partial J(x,t)}{\partial x} 
\end{equation}
where the current of probability is given by 
\begin{eqnarray}
J(x,t) &=&  \left[-F(x) + (1-\alpha) \frac{1}{2} D \frac{\partial \phi(x,p)}{\partial x}+	 \right. \nonumber \\
&&+ \left. \frac{1}{2}D\phi(x,p) \frac{\partial}{\partial x} \right]p(x,t).
\label{J} 
\end{eqnarray}
Here we have indicated that $\phi(x,p(x,t))$ could be a function of $p(x,t)$ given by Eq.~(\ref{phi}).

The equilibrium distribution is defined as the stationary solution with zero
current of probability, {\em i.e.}, 
\begin{equation}
P_{eq}(x)=\lim_{t\to\infty} p(x,t)
\end{equation}
supplemented with  $\lim_{x \to \pm \infty}J(x,t)=0$.
In the following sections we will find the equilibrium probability distribution
in the whole parameter range $\{\eta, \theta\}$.

\section{The linear case, $\eta=0$}
\label{eta0}
Let us begin by analyzing stationary states of the simpler case $\eta=0$. For
this case,
\begin{equation}
\phi(x)=D[g(x)]^\theta
\end{equation}
and the Fokker-Planck  equation (Eq.~(\ref{FP})) is linear. 
The stationary states have been studied in Refs.~\cite{Celia,Coutinho,ArenasBarci3} for different particular cases.
In this section we summarize the main results and procedures in order to present them in a unified scheme and to compare 
them with the nonlinear case. 

The equilibrium solution takes the form~\cite{ArenasBarci3} 
\begin{equation}
P_{eq}(x)=N\; e^{-U_{eq}(x)},  
\label{Peq}
\end{equation}
where $N$ is a normalization constant and the ``effective potential'' is given by  
\begin{equation}
U_{\rm eq}(x)=-2 \int^x \frac{F(x')}{D g(x')^\theta} dx'+ (1-\alpha) \theta\ln
g(x).
\label{Ueq}
\end{equation}
Thus, as already mentioned, the equilibrium distribution depends on the particular stochastic prescription used to define the 
Langevin equation.  
For general functions $F(x)$ and $g(x)$, the probability density distribution is given by Eqs.~(\ref{Peq}) and 
(\ref{Ueq}). The only constraint is a condition of integrability in order to
compute the normalization factor $N$.
To go further,  we need to impose constitutive relations between  drift and 
dissipation. 
For instance, suppose that the  system is submitted to a conservative force, with energy potential $V(x)$. For 
$\theta=0$, the resulting process is additive  and  
\begin{equation}
U_{\rm eq}(x)= \left(\frac{2}{D}\right)\;   V(x)
\label{Ueq0}
\end{equation}
up to an unimportant constant term that is absorbed in the normalization.
Then, Einstein relation imposes for the inverse temperature $\beta=2/D$, leading
to the Boltzmann distribution. 
On the other hand,  we could  impose, for $\theta\neq 0$,  a local generalization of Einstein relation 
\begin{equation}
F(x)=-\left(\frac{\beta}{2D g^{\theta}(x)}\right) \; V'(x) 
\end{equation}
ending with the solution~\cite{ArenasBarci2,ArenasBarci3}
\begin{equation}
U_{\rm eq}(x)=\beta V(x)+ (1-\alpha)\theta\ln g(x).
\label{Ueq1}
\end{equation}
We see that, for multiplicative noise, the final distribution is generally not of the Boltzmann type, even for the usual prescriptions of  It\^o ($\alpha=0$) or Stratonovich ($\alpha=1/2$). The exception is the anti-It\^o interpretation ($\alpha=1$) which, together with the local Einstein relation, leads to the usual thermodynamical equilibrium 
distribution. For this reason, this convention is also  called the kinetic prescription. 
An interesting particular case is to consider  a ``free'' particle in an inhomogeneous dissipative medium, where $V=0$ and the 
probability distribution is a power law of the form 
\begin{equation}
 P_{eq}= N \frac{1}{g(x)^{\theta(1-\alpha)}}\,,
\end{equation}
assuming it is normalizable.

Moreover, we could impose constitutive relations  different from the local Einstein relation, such as  the one used 
in Ref.~\cite{Celia}. We can choose, for instance,  
$F(x)=- V'(x)$,  $D g(x)= A + B V(x)$ and $\theta=1$; for simplicity we shall assume $A>0$ and $B>0$. Substituting these expressions 
in Eq.~(\ref{Ueq}) we immediately find 
a $q$-exponential form (Eq.~(\ref{q-exp})) with 
\begin{equation}
q =\frac{2(B+1)-\alpha B}{B +2 -\alpha B} \mbox{~~~~and~~~~} \beta = \frac{B(1-\alpha)+2}{A}. 
\label{qalpha}
\end{equation}

Therefore, we have shown that,  using the general solution Eqs.~(\ref{Peq}) and
(\ref{Ueq}), in the linear case $\eta=0$, we can find different types of equilibrium distributions, such as the Boltzmann or the $q$-exponential distribution, depending on the constitutive relation imposed between drift and dissipation and on the particular stochastic prescription used to derived the FP equation. 
Let us also notice that, whenever $A$ and $B$ have the same sign, the inverse temperature $\beta$ is positive, as normally expected; if both are positive (negative), then $q>1$ ($q<1$), which corresponds to long-tailed distributions (compact support distributions).

\section{The nonlinear case, $\eta\neq 0$}
\label{etaneq0}

The solution of the nonlinear and inhomogeneous FP equation (\ref{FP}) with (\ref{phi}) 
for general values of $g(x)$ and $F(x)$ is quite involved. We will look 
for solutions imposing the constitutive relations~\cite{MarizTsallis}
\begin{equation}
F(x)=-V'(x)\mbox{~~~~and~~~~}g(x)=A+BV(x),
\label{constraints}
\end{equation}
where, as already mentioned,  $A, B$ are real positive constants.

Looking for  stationary solutions $\partial p(x,t) / \partial t = 0$ and assuming appropriate boundary conditions 
which guarantee a null net flux, we have 
\begin{eqnarray}
\lefteqn{
\frac{\partial F(x)p(x,\infty)}{\partial x} = } \nonumber \\
&& \frac{D}{2}\frac{\partial}{\partial x}\left[ (1-\alpha)\frac{\partial \phi(x,p)}{\partial x} + \phi(x,p) \frac{\partial}{\partial x} \right]p(x,\infty).
\label{StationaryFP}
\end{eqnarray}
The choices made in Eq.~(\ref{constraints}) allow us to write the differential equation Eq.~(\ref{StationaryFP}) in terms 
of the variable $V$, obtaining
\begin{equation}
(1-\alpha)\frac{\partial [g(V)^{\theta}p(V)^{\eta}]}{\partial V}p(V) + g(V)^{\theta}p(V)^{\eta} \frac{\partial p(V)}{\partial V} = - \frac{2p(V)}{D},
\end{equation}
where $p(V)\equiv p(V(x),\infty)$. 
This equation can be re-written in the form of a \emph{Bernoulli equation}~\cite{Bernoulli} (see also \cite{bemski}),
\begin{eqnarray}
\frac{d p(V)}{d V}& +&
\frac{(1-\alpha)\theta B g(V)^{-1}}{[(1-\alpha)\eta +1]}p(V) = 
 \nonumber \\ 
 &&- \frac{2g(V)^{-\theta}}{D[(1-\alpha)\eta +1]}[p(V)]^{1-\eta},
\label{Bernoulli}
\end{eqnarray}
a class of nonlinear differential equations that can be linearized by a suitable change of variables. 

For $\eta=0$, we recover  the linear equation  we have treated in the last subsection. 
For $\eta\neq 0$ we can perform 
the  nonlinear change of variables, 
\begin{equation}
Z(V)=C_N\;  p(V)^\eta\; ,  \label{Z}
\end{equation} 
where $C_N$ is a normalization constant. With this, Eq.~(\ref{Bernoulli})
becomes a first order linear ordinary differential equation
\begin{equation}
\frac{d Z}{dV} + \frac{(1-\alpha)\eta \theta B g(V)^{-1}}{[(1-\alpha)\eta +1]}Z = - \frac{2g(V)^{-\theta}C_N \eta}{D[(1-\alpha)\eta +1]},
\label{ODE}
\end{equation}
with the general solution
\begin{eqnarray}
Z&=& (A+BV)^{-\frac{(1-\alpha)\eta \theta}{(1-\alpha)\eta +1}}
\left\{C_I - \frac{2 C_N \eta}{BD[(1-\alpha)\eta + 1- \theta]} 
\times \right. \nonumber \\
&\times& \left. \left[(A+BV)^{\frac{(1-\alpha)\eta +1 -\theta}{(1-\alpha)\eta +1}} - A^{\frac{(1-\alpha)\eta +1 -\theta}{(1-\alpha)\eta +1}} \right] \right\},
\label{Bernoulli_sol}
\end{eqnarray}
where $C_I$ is an integration constant which depends on the ``initial'' condition.
Thus, Eqs.~(\ref{Z}) and (\ref{Bernoulli_sol}) provides a family of explicit solutions of the nonlinear FP equation 
in terms of two constants, $C_N$ and $C_I$, which should be adjusted by means of  an initial condition and the probability distribution normalization.

\subsection{$\theta=0$}
In the particular  case $\theta= 0$, the inhomogeneity of the dissipation function $\phi(x)$ comes only from 
the probability density. The stationary solution can be read from Eq.~(\ref{Bernoulli_sol})
\begin{equation}
 Z=C_N p(V)^\eta=C_I\left(1- \frac{2C_N\eta}{D C_I[1+\eta-\alpha\eta]}\; V \right),
\end{equation}
which can be rewritten in terms of a $q$-exponential (Eq.~(\ref{q-exp})) with $N_q=(C_I/C_N)^{1/\eta}$, 
\begin{equation}
q=1-\eta \mbox{~~~~and~~~~}N_q^{1-q}\beta=\frac{2}{D(1+\eta-\alpha\eta)} \,.
\label{theta0}
\end{equation}
Interestingly enough, the index $q$ is $\alpha$-independent. We will show that this is a general feature of 
nonlinearity ($\eta\neq 0$). In contrast, the inverse temperature $\beta$ depends on the prescription that has been used. 
In particular, Eq.~(\ref{theta0}), in the It\^o prescription, i.e. $\alpha=0$, coincides with the result for the homogeneous 
nonlinear model obtained in Refs.~\cite{Plastino95,Tsallis96}.

\subsection{$\theta\neq0$}
For the inhomogeneous and nonlinear case,  i.e. $\theta \neq 0$, we can straightforwardly verify
\begin{equation}
C_I = -\frac{2C_N \eta  A^{\frac{(1-\alpha)\eta +1 -\theta}{(1-\alpha)\eta +1}}}{BD[(1-\alpha)\eta +1 -\theta]} \,.
\end{equation}
Consequently, using Eq.~(\ref{Bernoulli_sol}), we obtain
\begin{equation}
Z = - \frac{2C_N \eta}{BD[(1-\alpha)\eta +1 -\theta]}(A+BV)^{1-\theta} \,,
\end{equation}
hence, from Eq.~(\ref{Z}), 
\begin{equation}
p(V) = \left\{ -\frac{2 \eta}{BD[(1-\alpha)\eta +1 -\theta]} \right\}^{\frac{1}{\eta}}\!\!(A+BV)^{\frac{1-\theta}{\eta}} \,.
\label{p_sol}
\end{equation}
This expression can be written in the form of a  $q$-exponential (Eq.~(\ref{q-exp}))
with 
\begin{equation}
q = 1-\frac{\eta}{1-\theta},
\label{q}
\end{equation}
and the parameter $\beta$
\begin{eqnarray}
\beta &=& \frac{B}{A}\frac{1-\theta}{\eta} \\
&=& \frac{C_N}{C_I}\frac{2 (1-\theta)}{D[(1-\alpha)\eta + 1 -\theta]}A^{-\theta / (1-\alpha)\eta +1}.
\label{beta}
\end{eqnarray}

We see that the $q$-exponential distribution is a solution of the nonlinear and inhomogeneous FP equation {\em for any value of the stochastic prescription $\alpha$}. Moreover, the value of $q$ itself is {\em universal} in the sense that it does not depend on the prescription $\alpha$, as can be seen in Eq.~(\ref{q}). The index $q$ only depends on the parameters $\eta$ and $\theta$ that measure the relative importance of nonlinearity and inhomogeneity. 
Of course, this value of $q$ coincides with the one computed in Ref.~\cite{MarizTsallis} using the It\^o prescription.
Different stochastic prescriptions only affect the temperature parameter $\beta$. 

From Eq.~(\ref{q}), it could be wrongly concluded that in the linear limit $\eta\to 0$, the probability distribution is of the Boltzmann type, since $q\to 1$. However, as we showed in the previous section, in the linear case, we find power-law 
solutions with non-universal $q$ ($\alpha$-dependent). In other words, the solutions are {\em not} analytic in the linear 
limit $\eta\to 0$.  In variance with this fact, the limit $\theta\to 0$ is perfectly well defined as can be seen from Eqs.~(\ref{q}) and (\ref{theta0}).

\section{Conclusions}
\label{conclusions}
We have presented an inhomogeneous and nonlinear Fokker-Planck equation describing a generalized Markov stochastic process in 
the  ``generalized Stratonovich prescription''. This prescription is parametrized by a real parameter $0\leq \alpha \leq 1$, and contains the usual 
Stratonovich ($\alpha=1/2$), It\^o ($\alpha=0$) and kinetic ($\alpha=1$) prescriptions as particular cases. 
 We have also  parametrized nonlinearity and inhomogeneity by means of two parameters, $\eta$ and  $\theta$,
 in such a way  that the point $\eta=\theta=0$ represents a normal difusion process. In this way, we have
 unified and generalized  different results already obtained in 
the literature~\cite{Plastino95,Tsallis96,ArenasBarci3,MarizTsallis,Celia,Coutinho}.

\begin{table}
\footnotesize
\begin{tabular}{lclcl} 
\hline \hline
  Fokker-Planck  					&   &  ~~~~Linear                	&     & ~~~~Nonlinear	\\  
  equation				    	   &   &  ~~~~$(\eta = 0)$            &     & ~~~~$(\eta \neq 0)$ \\ [1ex]
\hline
Homogeneous					         &   &   Additive noise ~~~~    	&     & Multiplicative noise  \\
$(\theta = 0)$	            &   &     ~~~~ $q =1$          &     & ~~~~ $q = 1- \eta$  \\	
\hline
Inhomogeneous					       &   &   Multiplicative noise ~~  &     & Multiplicative noise  \\
$(\theta \neq 0)$	         &   & $q =\frac{2(B+1)-\alpha B}{B +2 -\alpha B}$ &   & ~~~~~$q = 1-\frac{\eta}{1-\theta}$ \\ [1ex]		
\hline
\end{tabular}
\caption{\footnotesize Stationary solutions of nonlinear inhomogeneous FP equation. The $q$-exponential distribution has been obtained for  the family of constitutive relations given by \eqref{constraints}. Notice that, in the nonlinear case, the exponent $q$  does not depend on the stochastic prescription, while it is not the case for the 
linear inhomogenous FP equation (the $q$ value for this case recovers the results in \cite{Celia} for the particular instances $\alpha=0$ and $\alpha=1/2$ by doing $B \to 2M/\tau C$; it also recovers, for $\alpha=0$, the results in \cite{MarizTsallis} by doing $B \to BD$).
Let us emphasize that $q$ cannot be arbitrarily large for a given $V(x)$, otherwise the normalizability property will be lost. For example, if we are dealing with $q$-Gaussians, then it must be $q<3$.}
\label{Summary}
\end{table}

We have solved the stationary FP equation for all values of the parameters $\eta, \theta, \alpha$.
For the linear case, $\eta=0$, we have found a general solution depending on the drift, the dissipation 
and the prescription $\alpha$. The Boltzmann distribution is obtained when a generalization of the Einstein relation and the kinetic prescription, $\alpha=1$, are imposed.  In all other cases, the solution is more involved.

There exist a link between the stationary solutions of the linear or nonlinear FP equation and the distribution obtained 
by extremizing a particular entropy under simple specific constraints. Indeed, the connection between the linear FP 
equation and Boltzmann-Gibbs entropy is well known since long. Analogously, for nonlinear FP equations yielding
specific classes of anomalous diffusion, nonadditive entropic functionals have been 
analyzed in detail~\cite{Plastino95,Tsallis96}. In addition, this remarkable link has also been found for 
even more general nonlinear FP equations and entropic forms~\cite{CuradoNobreetal,RibeiroTsallis}.

We analyzed a family of  constitutive relations between drift and dissipation that results in a $q$-exponential distribution, thus exhibiting a possible mechanism compatible with nonextensive statistical mechanics. In the nonlinear case, $\eta\neq 0$, the value of the exponent $q$ is, remarkably enough, $\alpha$-independent. The different prescriptions that define the stochastic process {\em only} affect the inverse temperature  $\beta$.
In the linear case, in contrast, the exponent of the power-law does depend on the stochastic 
prescription $\alpha$. This clearly shows that   the linear limit of the solutions is 
not analytic, namely, $  \lim_{\eta\to 0} p_{eq}[\eta,\theta]\neq p_{eq}[0,\theta] $.

Table \ref{Summary} summarizes the values of the entropic index $q$ for the $q$-exponential distributions obtained as the sta\-tio\-na\-ry-state distributions for nonlinear inhomogeneous Fokker-Planck equation when using the particular relations given by \eqref{constraints}. A variety of possible physical applications of the present results can be found in \cite{books} and references therein.

\acknowledgments

The Brazilian agencies CNPq and FAPERJ are acknowledged for partial  financial support. ZGA is a CNPq Postdoctoral Fellow.

\end{document}